Three-dimensional Ising models –Equations for Spontaneous Magnetization and Critical Temperature

M V Vismaya and M V Sangaranarayanan*

Department of Chemistry

Indian Institute of Technology Madras-Chennai-600036 India

E mail: vismayaviswanathan11@gmail.com; E mail: sangara@iitm.ac.in

Abstract

The equations for the spontaneous magnetization for different three-dimensional lattices have been derived in a heuristic manner. The estimated critical temperatures for simple cubic, face-centered cubic, body-centered cubic and diamond lattices are in excellent agreement with recent computer simulation data.

1. **Introduction**

   The analysis of two and three-dimensional Ising models continues to be a frontier area of research in statistical physics, in view of the absence of phase transitions in the one-dimensional analogues [1]. While Onsager's exact solution of two-dimensional Ising models for square and rectangular lattices provided an explicit expression for the partition function and other thermodynamic properties, the results are valid only when the external magnetic field is zero [2]. Furthermore, the complete derivation of the spontaneous magnetization $M_0$ in two dimensions is due to Yang [3] viz

$$M_0 = \left(1 - \frac{1}{sinh^4(2K)}\right)^{\frac{1}{8}} \tag{1}$$

with $K = \frac{J}{kT}$ and $J$ denotes the nearest neighbour interaction energy.

Here, we report new equations for the spontaneous magnetization for a few three-dimensional Ising models by exploiting the analogy with Onsager's exact solution for two-dimensions.

2. **Equation for the critical temperatures**

   Analogy with Onsager's solution for two dimensions

   In the Onsager's exact solution of two-dimensional Ising models for square lattices, the interaction energy parameter $\kappa$ defined as

$$\kappa(2d) = \frac{2\sinh(2K)}{cosh^2(2K)} \tag{2}$$

equals unity at the critical temperature, the precise value of $K$ at the critical temperature being $0.5\sinh^{-1}(1)$ for square lattices. It is imperative to investigate whether analogous equations for $\kappa$ and $M_0$ can be postulated for three dimensional Ising models too. In Table 1, we provide the equations for $\kappa(3d)$ which yield agreement with reported critical temperatures, for simple cubic, face-centered cubic, body-centered cubic and diamond lattices. The critical temperatures reported in Table 1 arises when the corresponding equations for κ are solved when $\kappa = 1$.

**Table 1: Critical temperatures for various three-dimensional lattices**

| Lattice | Equation for $\kappa$ (3d) | $\frac{J}{kT_c}$ using $\kappa(3d)$ | $\frac{J}{kT_c}$ from simulation |
|---|---|---|---|
| simple cubic | $\kappa(3d) = 2\frac{\sinh(3.9764\,K)}{\cosh^2(3.9764K)}$ | 0.221651 | 0.2216595[4] |
| face-centered cubic | $\kappa(3d) = 2\frac{\sinh(8.635\,K)}{\cosh^2(8.635K)}$ | 0.10207 | 0.102065[5] |
| body- centered cubic | $\kappa(3d) = 2\frac{\sinh(5.6005\,K)}{\cosh^2(5.6005\,K)}$ | 0.157374 | 0.157371[5] |
| diamond | $\kappa(3d) = 2\frac{\sinh(2.38385\,K)}{\cosh^2(2.38385\,K)}$ | 0.369727 | 0.369720 [5] |

The values for $\frac{J}{kT_c}$ reported in Table 1 are in complete agreement with those arising from simulation data [4,5]. Interestingly, the mathematical form of $\kappa$ given by Onsager in the exact solution of two dimensions is kept intact even for three dimensions, the two changes being (i)the replacement of the arguments of the hyperbolic functions and (ii) inclusion of new critical exponents. This may probably imply that the extension of Onsager's formalism to three-dimensional models is indeed feasible, *mutatis mutandis*. Fig 1 depicts the variation of κ with 'temperature' and 'inverse temperature' for the four lattices. It is of interest to note that this dependence is similar to that observed in the two-dimensional analogues. (cf. Fig 13.12 of [1].)

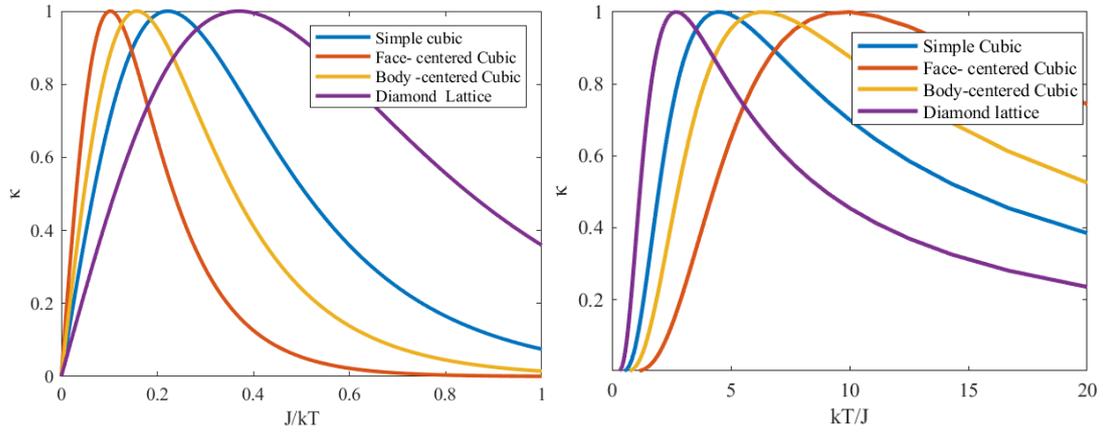

**Fig 1: Dependence of κ on (a) J/kT and (b) kT/J estimated using the equations shown in Table 1**

## 3. Equation for the Spontaneous magnetization

While the above eqns for $\kappa$ yields only the expression for critical temperatures, the corresponding eqns for the spontaneous magnetization is more involved. In the case of two-dimensions, the relation between $M_0$ and $\kappa$ is given by

$$\frac{\kappa}{2} = \frac{[1 - M_0^8]^{\frac{1}{4}}}{\left[1 + (1 - M_0^8)^{\frac{1}{2}}\right]}$$

In an analogous manner, a similar eqn can be *postulated* for $M_0$ in the case of three-dimensional Ising models using the eqns for κ shown in the Table 1, in conjunction with the critical exponent as 0.326423[6]. Table 2 provides the empirical equations for the spontaneous magnetization for various three-dimensional lattices.

**Table 2: Spontaneous magnetization equations for three-dimensional lattices**

| Lattice | Equation for the spontaneous magnetization |
|---|---|
| simple cubic | $M_0 = \left[1 - \dfrac{1}{\sinh^{1.531754}(3.9764\,K)}\right]^{0.326423}$ |
| face-centered cubic | $M_0 = \left[1 - \dfrac{1}{\sinh^{1.531754}(8.635\,K)}\right]^{0.326423}$ |
| body-centered cubic | $M_0 = \left[1 - \dfrac{1}{\sinh^{1.531754}(5.6005\,K)}\right]^{0.326423}$ |
| diamond | $M_0 = \left[1 - \dfrac{1}{\sinh^{1.531754}(2.38385\,K)}\right]^{0.326423}$ |

Figure 2 depicts the dependence of the spontaneous magnetization on 'temperature' and 'inverse temperature' , estimated using the equations shown in Table 2. This variation is similar to that observed in the case of two-dimensional models (cf. Fig 13.15 of [1])

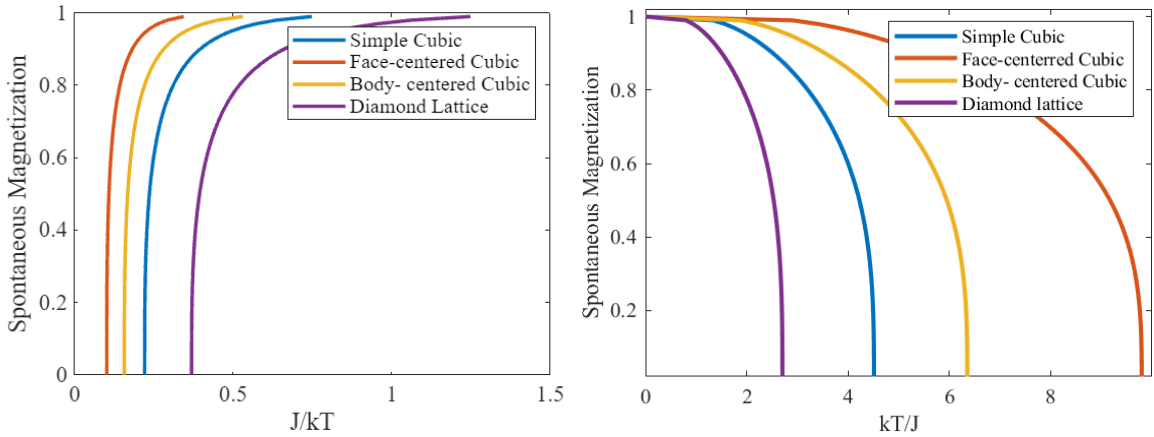

**Figure 2: Dependence of the spontaneous magnetization on (a) J/kT and (b) kT/J for various lattices, estimated using the equations in Table 2.**

4. **Results and Discussion**

It is well-known that the exact analysis of critical phenomena pertaining to three-dimensional Ising models is a long-standing problem in statistical physics. In view of the complexity of the problem, formulation of rigorous and correct solutions is rendered difficult [7]. A welcome feature in this context is the availability of series expansions at high and low temperatures. These have provided accurate estimates for the partition functions [8].

To deduce the critical parameters, several novel strategies have been employed. The precise values of the critical exponents have been of much debate in view of the accuracy and reliability of the simulation protocols employed therein. Among many estimates available for the critical parameters, the studies by Lundow et al [5], Fisher and Perk [6], Hasenbusch [9] are noteworthy. Sastre [10] has employed the probability distribution methods to estimate accurate values of the critical temperature and exponent for the spontaneous magnetization pertaining to simple cubic lattices.

At this stage, the equations provided herein are semi-empirical and guided by the mathematical structure of Onsager's equations for the spontaneous magnetization. Further studies are in progress to derive the spontaneous magnetization commencing from the corresponding partition functions.

## 5. Summary

The equations for the spontaneous magnetization have been derived for simple cubic, face-centered cubic, body-centered cubic and diamond lattices, in a heuristic manner. The estimated critical temperatures are in excellent agreement with recent computer simulation data.

**Acknowledgements**

This work was supported by the Mathematical Research Impact Centric Scheme (MATRICS) of Science and Engineering Research Board, Government of India. We thank the P.G. Senapathy center for computing resources, Indian Institute of Technology-Madras for computational facilities.